\documentclass[a4paper,fleqn,usenatbib]{mnras}

\usepackage[T1]{fontenc}
\usepackage{ae,aecompl}

\usepackage{graphicx}	
\usepackage{amsmath}	
\usepackage{amssymb}	
\usepackage{subfigure}
\usepackage{csquotes}

\title[The detectability of Wolf-Rayet Stars]{The detectability of Wolf-Rayet Stars in M33-like spirals up to 30\,Mpc} 

\author[J. L. Pledger et al.]{
J. L. Pledger,$^{1}$\thanks{E-mail: jpledger@uclan.ac.uk}
A. J. Sharp,$^{1}$
and A. E. Sansom, $^{1}$
\\
$^{1}$Jeremiah Horrocks Institute, University of Central Lancashire, Preston, PR1 2HE, UK \\
}

\date{Accepted XXX. Received YYY; in original form ZZZ}

\pubyear{2020}

\begin{document}
\label{firstpage}
\pagerange{\pageref{firstpage}--\pageref{lastpage}}
\maketitle

\begin{abstract}

We analyse the impact that spatial resolution has on the inferred numbers and types of Wolf-Rayet (WR) and other massive stars in external galaxies. Continuum and line images of the nearby galaxy M33 are increasingly blurred to mimic effects of different distances from 8.4\,Mpc to 30\,Mpc, for a constant level of seeing. We use differences in magnitudes between continuum and Helium II line images, plus visual inspection of images, to identify WR candidates via their ionized helium excess. The result is a surprisingly large decrease in the numbers of WR detections, with only 15\% of the known WR stars predicted to be detected at 30Mpc. The mixture of WR subtypes is also shown to vary significantly with increasing distance (poorer resolution), with cooler WN stars more easily detectable than other subtypes. We discuss how spatial clustering of different subtypes and line dilution could cause these differences and the implications for their ages, this will be useful for calibrating numbers of massive stars detected in current surveys. We investigate the ability of ELT/HARMONI to undertake WR surveys and show that by using adaptive optics at visible wavelengths even the faintest (M$_{V}$\,=\,--3\,mag) WR stars will be detectable out to 30\,Mpc.

\end{abstract}




\section{Introduction}

Evolutionary paths of evolved, massive stars are currently uncertain
and this affects our ability to understand which progenitors lead to
various categories of core-collapse supernovae (ccSNe)
\citep{Eldridge2013, Smith2011, Groh2013}. This study aims to quantify
the effects of spatial resolution on the detection of Wolf-Rayet (WR)
stars so that the numbers of massive stars can be correctly
interpreted with respect to stellar evolutionary theory.

Wolf--Rayet (WR) stars are evolved, core-He burning stars with initial
masses above $\sim$20M$_{\sun}$. Most of their hydrogen envelope has
been stripped due to violent stellar winds, with mass--loss rates of
$\sim$10$^{-4}$--10$^{-5}$M$_{\odot}$yr$^{-1}$
\citep{Nugis2000}. Prominent, broad emission lines arise as a result
of these strong winds with speeds of $v\geq$1000\,km~s$^{-1}$. The
metallicity dependence of these stellar winds means that more
metal-rich WR stars possess stronger winds and are able to strip even more of their outer envelopes \citep{Vink2005}. WR stars, down to lower masses, can also occur due to mass stripping in close binaries \citep{Gotberg2017, Shenar2019}, leading to a complexity of evolution. These close binaries may end their lives as merged black holes, high mass interacting binaries, or as asymmetric supernova explosions, where the black hole gets a kick large enough to prevent a high mass x-ray binary (HMXB) from forming \citep{Vanbeveren2020}.

As the WR stars undergo stripping they exhibit the products of
core-hydrogen burning, namely, helium and nitrogen in their
emission-line spectra and are classed as WN stars. More massive, or
more metal--rich WR stars undergo enhanced stripping revealing carbon
and oxygen produced during core-helium burning and are classed as
WC stars \citep{Conti1976}. WR stars can be split into ``early'' and ``late'' subtypes determined
by emission line ratios which represent the temperature of the
star. The spectra of cooler, late--type WNL and WCL stars are dominated
by stronger N{\sc iii} and C{\sc iii}, respectively, compared to N{\sc
  iv} and C{\sc iv} for hotter, early-type WNE and WCE stars,
respectively (see \citealt{Crowther2007} for a full review of WR
classifications).

Current stellar evolutionary models predict that the WC/WN and WR/O
ratios increase with metallicity as a result of metal--driven stellar
winds during the WR and O star phase \citep{Eldridge2006,
  Meynet2005}. Theory is currently in disagreement with observational
evidence, particularly at higher metallicity where even fewer WN stars
are detected than predicted, however \citet{Neugent2011}(hereafter
NM11) argue that this could simply be a result of WN stars being
harder to detect than WC stars. Indeed, when we look at individual WR stars in the Large Magellanic
Cloud (LMC) (\cite{Bibby2010}, their Figure 11) we see that WC stars have the
strongest emission lines and when observed through a narrow--band He\,{\sc ii} filter may appear over 2 magnitudes brighter than when viewed through a continuum filter; we refer to this as emission-line excess. WNE stars have a slightly weaker emission lines excess than WC stars, with the star appearing $\sim$0.5--2\,mag brighter in the He\,{\sc ii} image compared to a continuum image. WNL stars have the weakest emission line excess of $<$1\,mag and this creates a natural bias towards detecting WC stars in WR surveys.

However, even in galaxy surveys within the Local Group we are unlikely
to be resolving individual WR stars. In large, dense regions the overall brightness is increased but the emission-line excess can be decreased as a result of emission line dilution. For example, in the LMC, based on a spatial scale of $\sim$0.25\,pc the R136 cluster has an (unresolved) synthetic narrow-band magnitude of M$_{4686}$ $\sim$ --10\,mag \citep{Bibby2010}
but an excess of only m$_{4781}$--m$_{4686}$\,=\,0.15\,mag. \citet{MasseyHunter1998} used HST spectroscopy to identify 65 of the bluest, hottest stars, including several H-rich WN stars present in R136.
\citet{Doran2013} used the VLT-FLAMES Tarantula Survey data to carry out a census of the 30 Doradus star forming region in the LMC, including R136 star cluster, and found that the inner 5\,pc hosts 12 WR stars and $\sim$70 O stars. More recently, \citet{Crowther2016} presented new HST data from the R136 central star cluster, classifying the massive stars. Of the 51 stars they obtained HST/STIS spectroscopy of, 24 of them (three WN5, 19 O-type dwarfs and two O-type supergiants) lie within 0.25\,pc (1 arcsecond) of R136a1 (\citealt{Crowther2016}; their Table 4). This spatial resolution is typical of resolutions achieved with ground-based observations where adaptive optics is not available.

It is the continuum emission from these other blue stars within the same (unresolved) region as the WR stars that causes the strength of the He\,{\sc ii}$\lambda$4686, N\,{\sc iii}$\lambda$4630 and C\,{\sc iii}$\lambda$4650 emission lines to be diluted. For nearby galaxies such as the LMC (50kpc) and M33 (0.84Mpc), where we can resolve sources down to parsec (or sub-pc) scales, the
problem of emission line dilution is minimal. Beyond the Local Group this dilution can
significantly reduce the emission line excess and severely impact our
ability to detect WR stars. A simple, proof-of-concept test by \citet{Bibby2010} estimated that if
the LMC stars were located at a distance of $\sim$4\,Mpc, and the
spatial resolution went from 0.25\,pc to 25\,pc (assuming a ground-based 1
arcsecond seeing), then 20\% of the stars would not be detected as a
result of line dilution. A full study to quantify the detectability of
WR stars, including line dilution, has not been undertaken and this is our aim with the work presented in this paper. 

The detectability of WR stars is important not only in the context of
stellar evolutionary models but also for identifying the progenitors
of Type Ibc ccSNe. \textbf{Single} WN and WC stars have
been predicted to be the progenitors of H-poor and H+He--poor Type Ib
and Ic ccSNe \citep{Ensman1988}, however binary star systems can also produce WR stars as a result of Roche Lobe Overflow and common-envelope evolution which would also produce a Type Ib/c ccSNe. \citep{Gotberg2017, Eldridge2017, Vanbeveren1998}. Moreover, current stellar evolutionary models suggest that the majority of Type Ib/c ccSNe can be produced from close binary stars of much lower initial mass \citep{Dessart2020}.

Over the past two decades attempts have been
made to use archival, pre-SN imaging to identify progenitors and confirm a single or binary evolutionary scenario. This has been very successful for hydrogen--rich Type II SN (See
\citealt{Smartt2009} for a review), but less so for Type Ibc SNe (See \citealt{Eldridge2013} and \citealt{VanDyk2017} for a review).

Type Ib SN iPTF13bvn in NGC 5806 at $\sim$22\,Mpc is one example. \citet{Cao2013} detected a bright source in archival imaging but concluded that it was
too bright for a single WR star. A binary progenitor was favoured by
\citet{Bersten2014} based on modelling of the light curve, suggesting
that at least some Type Ib SNe result from a binary evolutionary
channel. Post-SN HST imaging shows that the progenitor has disappeared
\citep{Eldridge2016} and predicts an initial mass of 10-12M$_{\odot}$
but deep UV observations presented in \citet{Folatelli2016} suggest that the majority of the flux in the pre--SN images came from the SN progenitor itself rather than a binary companion. Most recently, \citet{Kilpatrick2021} combined pre- and post-explosion imaging to identify a progenitor candidate for SN2019yvr that may have experienced significant mass-loss prior to explosion. The cool temperature of the progenitor is inconsistent with the lack of hydrogen in the SN spectra and it remains unclear if the progenitor is a single or binary star. 

Progenitors of Type Ic SN remain elusive, with only SN2002ap
\citep{Crockett2007} having a detection limit deep enough to rule out
a WR star and support a binary scenario. Similarly, detection limits
of post-SN HST imaging of Type Ic SN1994I rule out any binary
companion $>$10M$_{\odot}$ \citep{VanDyk2016}. Such observations were
only possible due to the availability of HST imaging and the
relatively near distance of the host galaxy, M51 at $\sim$8Mpc. Most recently, Pre--SN HST imaging of SN2017ein in NGC 3938, at an (uncertain) distance of 17--22\,Mpc, identified a luminous, blue progenitor candidate which could be either a single star, binary system or star cluster \citep{VanDyk2018} based on a spatial resolution of 12--16\,pc. 

High resolution imaging is essential if we are to resolve the single versus binary progenitor debate for ccSNe; this is the motivation for this work. In this paper we use observations of M33 to investigate the effect that emission line
dilution has on the detectability of WR stars. The most distant
supernova progenitor detected is the LBV progenitor of Type IIn SN
2005gl at 66\,Mpc \citep{VanDyk2017}. However, at these distances it
becomes extremely difficult to identify single massive stars and to
have a reasonably complete supernova detection rate, thus we set our
upper limit at 30Mpc.

In Section \ref{M33} we review the WR content of M33. In Section \ref{method} we give the details of our method and analysis followed by the results in Section \ref{results}. Section \ref{discussion} puts our results into context in terms of the predictions from stellar evolutionary models and the implications for detecting supernova progenitors up to 30\,Mpc. 

\section{The Wolf-Rayet Content of M33} \label{M33}

M33 (the Triangulum Galaxy) is a face-on, SA(s)cd spiral galaxy at a
distance of just 839\,kpc\citep{Gieren2013}, and thus is well studied in
the literature. Some WR surveys have targeted the giant H\,{\sc ii}
regions of M33, such as NGC 604 \citep{Bruhweiler2003}, NGC 595 and NGC
592 \citep{Drissen2008} however, the first complete WR survey of the
full galaxy was undertaken by NM11 (including earlier
work by \citet{Massey1998}), and identified 206 WR Stars.

In the NM11 survey WR candidates were identified using
images taken with the Mosaic CCD camera on the Kitt Peak 4.0m Mayall
telescope through three narrow--band filters; `WR He\,{\sc ii}'
($\lambda_C$=4686\AA), `WR C\,{\sc iii}' ($\lambda_C$=4650\AA) and
a continuum filter `WR 475' ($\lambda_C$=4750\AA). Three fields
(each covering 36$\arcmin$ $\times$36$\arcmin$) were obtained per
filter to cover the extent of the galaxy's spiral arms and the
telescope was dithered between these exposures to fill the chip
gaps. The seeing in the central and southern fields was 1.1$\arcsec$
increasing to 1.5$\arcsec$ in the northern field. WR candidates were
identified through both photometry and visual inspection of the
continuum subtracted He\,{\sc ii} and C\,{\sc iii} images. The
3-filter approach used in these observations allowed
NM11 to not only identify WR candidates but also to
predict their WC or WN subtype. This was confirmed via follow-up
spectroscopy obtained from the Hectospec instrument on the 6.5m
Multiple Mirror Telescope.

This catalogue of WR stars was updated by \citet{Neugent2014} to add
six new WN stars and to declassify an existing object in the 2011
catalogue (an LBV originally identified as a B0.5Ia$+$WNE system),
taking the total known WR population to 211. The final catalogue consists of
148 WN stars, 52 WC stars, 9 Ofpe/WN9 stars (also known as WN9-11 stars \citep{Crowther1995, Crowther1997}) and two WN/WC transition stars. NM11 consider their survey to be complete to 95\%.

\section{Method} \label{method}

The images of M33 from NM11 were kindly provided to us
fully reduced by Philip Massey. They covered M33 with three pointings
and obtained three, 300 second exposures of each field in the He\,{\sc
  ii}~$\lambda$4686, C\,{\sc iii}$~\lambda$4650 and continuum~$\lambda$4750 filters. 

NM11 did not stack these multiple exposures
in order to preserve the best photometry possible \citep{Massey2006}. However,
to detect the faintest WR stars in more distant galaxies longer
exposure times (of order 3000 seconds) are required, therefore multiple exposures must be
stacked. In order to use a consistent approach across all distances we combined the images from NM11, using the \textsc{imcombine} routine in \textsc{iraf} \citep{Tody1986}, as done for WR surveys of more distant galaxies. 

\begin{figure}
\includegraphics[trim={2cm 7cm 1cm 4cm}, clip, width=0.92\columnwidth, angle=90]{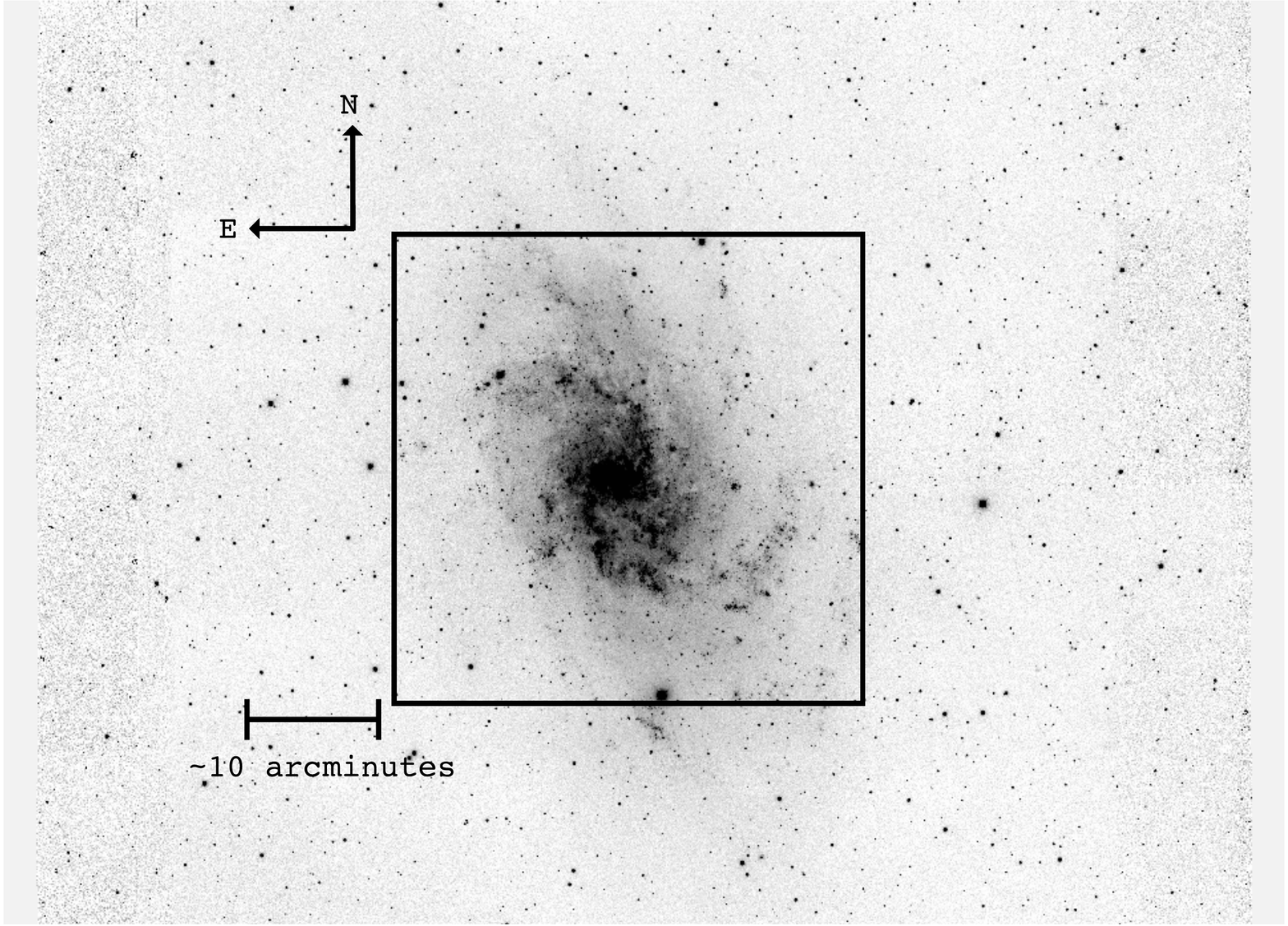}
\caption{Moses Holden Telescope B-band image of the full field of M33
  (Credit: M.Norris), showing the field of view covered by the central
  36$\times$36 arcminute pointing of the Kitt Peak He\,{\sc ii} data
  from NM11.}
\label{M33_fov}
\end{figure}

For galaxies beyond the Local Group, 8--m class telescopes are required in order to reach the magnitude limit for WR stars (typically M$_{V}$\,=\,--3\,mag \citep{Groh2013b}). However,  C\,{\sc iii} narrow-band filters are not widely available at these facilities so we chose to only
use the He\,{\sc ii} and continuum imaging, not the C\,{\sc iii}
images. NM11 used three (central, northern and
southern) pointings to cover the full field of M33 and identify the
211 WR stars but in practice with large facilities one pointing is
more achievable. We concentrate our study on the central pointing of
M33, the area of which is illustrated in Figure \ref{M33_fov}, using a larger scale
image taken with the Moses Holden Telescope at UCLan, Preston. This
central region contains 196 (93\%) of the WR stars.

\subsection{Image Degradation}

To investigate the detectability of WR stars out to 30\,Mpc we
artificially degrade the original M33 image to mimic the observations
typical of star-forming spiral galaxies at different distances; this was
done using the \textsc{iraf} routines \textsc{gauss} and
\textsc{blkavg}.

The \textsc{gauss} routine convolves the image with a specified
Gaussian profile to blur the image. The input width required for the
convolution was calculated by using an equation derived from the
convolution of two Gaussian functions, which resulted in the formula
$\sigma_g$=$\sigma_{M33}$$\sqrt{\delta^2 - 1}$ where $\sigma_{M33}$ is
the full width at half maximum (FWHM) of the stars in the original
non-degraded images and $\delta$ /= D$_{\deg}$/D$_{M33}$ where D$_{deg}$
is the distance to which the image was to be degraded to, and
D$_{M33}$ is the distance to M33.

The \textsc{blkavg} routine takes x $\times$ y integer pixels in the
initial image and bins them to just one pixel in the output image,
which takes a value of the mean of the x $\times$ y pixel values in
the initial image. In this study, the images were binned by factors of
2$\times$2, 5$\times$5, 10$\times$10, 20$\times$20 and 36$\times$36,
representing galaxy distances of 1.68Mpc, 4.20Mpc, 8.39Mpc, 16.78Mpc
and 30.2Mpc, respectively for ground-based observations with
1.1\arcsec~seeing. The original image can resolve sources to a FWHM\,=\,4.5\,pc, with the subsequent distances representing spatial resolutions
of 9\,pc, 22.4\,pc, 44.8\,pc, 89.6\,pc and 161.1\,pc, respectively.

Figure \ref{blink} shows the image degradation for two sources in the
He\,{\sc ii} (top), continuum (middle) and continuum subtracted
(bottom) images. The central source in the bottom image is a WC4 star
and the star to the lower right is a WN6($+$abs) star. Both sources
start to blend into a single source, (along with other stars) at
16.8Mpc.

\subsection{Photometry}

The images of M33, blurred to different spatial resolutions were then
analysed to identify the WR population. WR stars exhibit He\,{\sc ii} emission lines which are stronger than the continuum emission, hence for a comparison of He\,{\sc ii} and continuum filter photometry we expect WR sources to have $\Delta$m<0 i.e. the star emits more flux in the He\,{\sc ii} filter than the continuum filter; this is known as the He\,{\sc ii} excess.

The \textsc{daophot} package
\citep{Stetson1990} within \textsc{iraf} was used to carry out point
spread function (PSF) crowded field photometry on the combined M33
images. Photometry was performed on both the He\,{\sc ii}~$\lambda$4686 and the Continuum $\lambda$4750 images, then sources were matched in terms of their x,y co-ordinates and the magnitude difference $\Delta$m\,=\,m$_{4686}$--m$_{4750}$ calculated for
each source. WR sources with $\Delta$m<0 are identified and only accepted as a WR candidate if the He\,{\sc ii} excess is significant at $\geq$3$\sigma$ level, which we determine using the associated error on the photometry. 

\subsection{Inspection of the Image}
\begin{figure}
\centering
\includegraphics[width=0.98\columnwidth, trim={2cm 13cm 2cm 0cm}, clip]{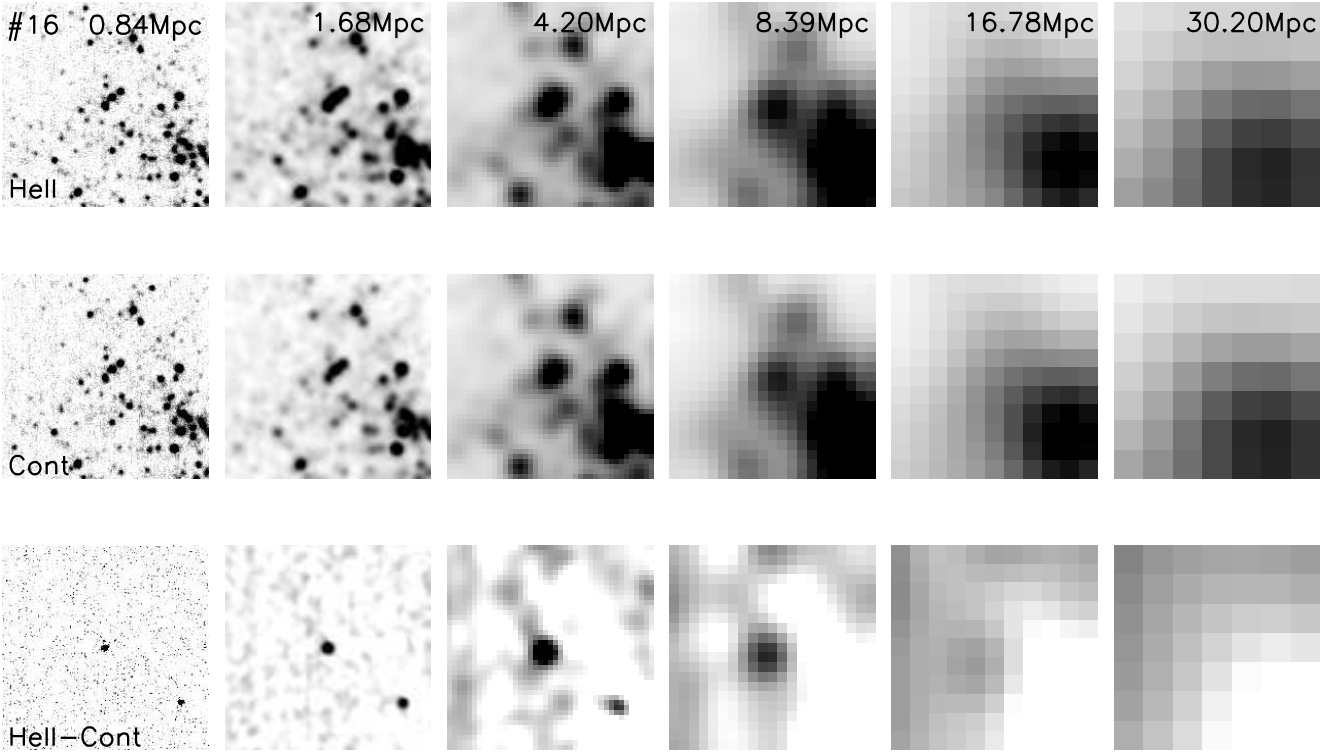}
\caption{The central source, Source \#16, is a WC4 star detected in
  the He\,{\sc ii} filter (top), the Continuum $\lambda$4750 filter
  (middle) and the continuum subtracted image (bottom). The source in
  the lower right of the continuum subtracted image is a WN6 star. The
  image is degraded to represent the spatial resolution achievable at
  increasing distances, under the same seeing conditions. The box size
  corresponds to 1 arcsecond diameter at each distance. It is clear
  that even at 4\,Mpc sources merge making it harder to identify the WR
  stars. The continuum-subtracted image reveals the WC star at 0.84\,Mpc
  without any confusion of surrounding sources and the unresolved star can still be
  detected in all images within 8.4Mpc. However, at 16.8\,Mpc the star cannot be seen in the He,{\sc ii} or continuum filters but can be detected in the continuum-subtracted image, but at 30.2\,Mpc the WR stars is not detected at all.}
\label{blink}
\end{figure}

An additional method for identifying WR candidates in narrow-band
imaging is to look for sources with helium excess in the continuum
subtracted image by \enquote{blinking} (\citealt{Moffat1983, Massey1983}) which is demonstrated in
Figure \ref{blink}. 

Analysis of the continuum subtracted images is a useful tool for a
number of reasons. Firstly, it can help to remove false positives
that have no visible helium excess, even though they were determined as having a He\,{\sc ii} excess from photometry. This is reflected in
NM11 and \citet{Bibby2010} who choose a $\Delta$m cut
of 0.1\,mag and 0.15\,mag, respectively. Secondly, it can show sources
of helium excess that the photometry failed to pick up, for example in
crowded regions. Finally, the continuum subtracted image can identify
WR candidates which are only detected in the He\,{\sc ii} image and
not the continuum; these candidates would not be picked up via
photometry.

\onecolumn

\twocolumn

\vspace{-0.25cm}
\section{Results} \label{results}
We have performed the analysis outlined in Section \ref{method} on the
images at 0.84\,Mpc, 1.68\,Mpc, 4.195\,Mpc, 8.39\,Mpc, 16.78\,Mpc and 30.2\,Mpc to
quantify the detectability of WR stars at the corresponding spatial
resolutions. A summary of our results is presented in Table
\ref{detections}. At the true distance of M33, 0.84\,Mpc, we detect 93\%
of the WR stars detected by NM11. Using the a priori information from NM11 the missing WR stars
were identified as bad subtractions. Investigating this further, we found that many of the WR stars were identified by our photometry with a small He\,{\sc ii} excess
($<$0.3\,mag) but they were ruled out as WR candidates after inspection of
our continuum subtracted image. These bad subtractions are a result of
the PSF being compromised by the combining of our images and is why NM11 chose not to stack their images. At our furthest distance of 30\,Mpc, or 161\,pc
resolution, we only detect 15\% of the known WR stars in the central pointing of M33, as shown in Figure \ref{M33_fov}.

At increasing distances the WR stars can either (i) blend into another
source and the emission line is diluted until the WR star
is no longer detected (ii) fade into the background noise until it is
no longer detectable or (iii) blend with additional WR stars so the
emission line excess is increased. In practice, both (i) and (iii) can
occur together at some level depending on the distribution of WR
stars.

\begin{table}
\caption{Results of our analysis at 6 distances. 93\% of the 196 known WR stars in the central pointing are detected in the undegraded (but combined) image decreasing to 15\% at
  30\,Mpc. N(WR)$_{phot}$ and N(WR)$_{net}$ are the number of WR stars
  detected via photometry and additionally from the
  continuum-subtracted image, respectively. We note that a
  $\geq$3$\sigma$ detection was required for the source to be classed
  as a photometric detection.}
\begin{tabular}{c@{\hspace{2mm}}c@{\hspace{2mm}}c@{\hspace{2mm}}c@{\hspace{2mm}}c}
\hline
Distance  & Spatial  & N(WR)$_{phot}$ & N(WR)$_{net}$ & N(WR)$_{total}$ \\
(Mpc) & Resolution  & \\
 & (pc) \\
\hline
0.84  & 4.5    & 172 & 11 & 183 (93\%) \\
1.68  & 9.0    & 111 & 47 & 158 (81\%) \\
4.20  & 22.4   & 56  & 76 & 132 (67\%) \\
8.39  & 44.8   & 2   & 74 & 76 (39\%) \\
16.78 & 89.6   & 13  & 35 & 48 (24\%) \\
30.20 & 161.1  & 12  & 18 & 30 (15\%) \\
\end{tabular}
\label{detections}
\end{table}

\vspace{-0.25cm}
\subsection{Emission Line Excess}

\begin{figure}

\vspace{-0.5cm}
\subfigure{\includegraphics[width=0.5\columnwidth]{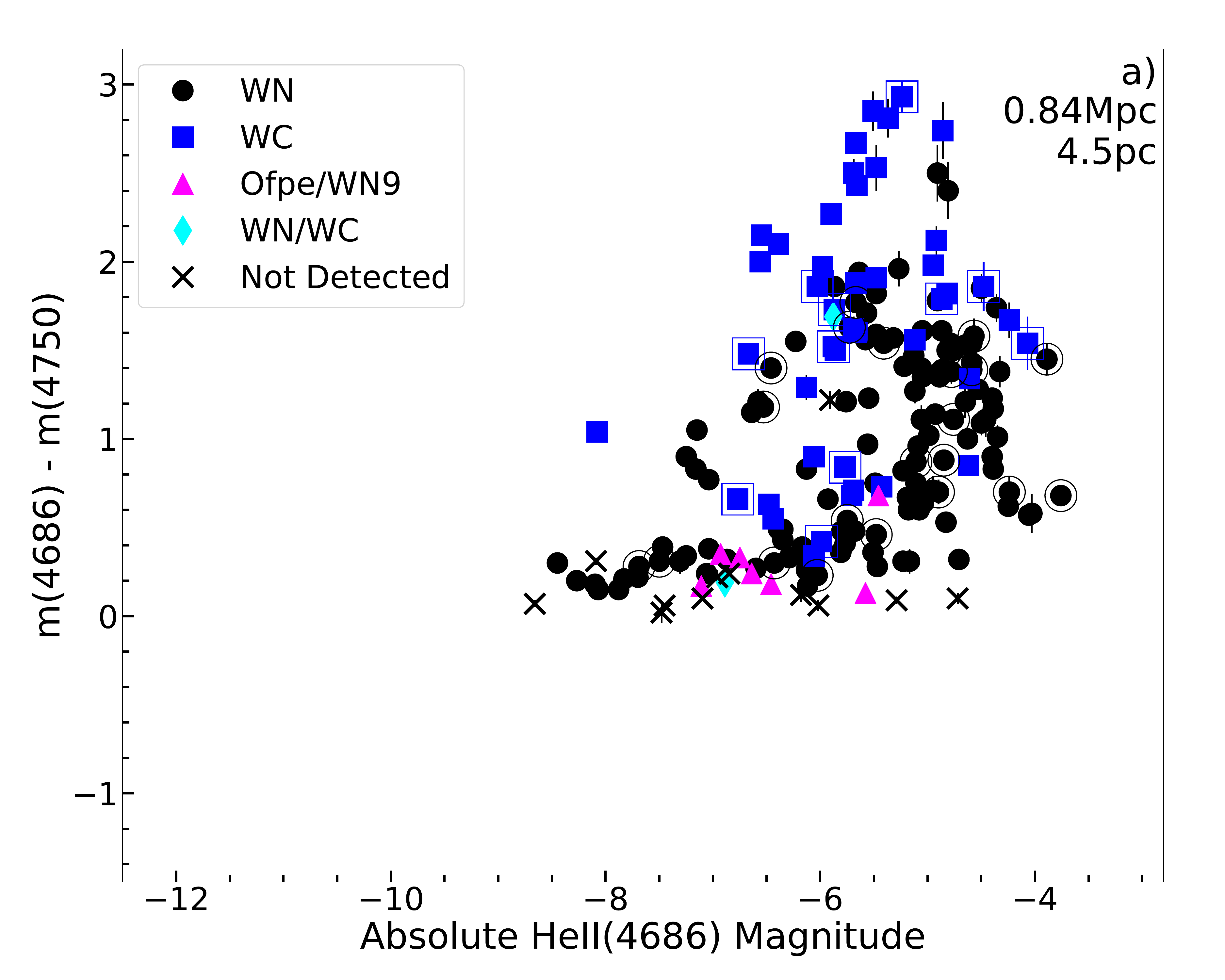} \label{0_deg}}
\vspace{-0.5cm}
\subfigure{\includegraphics[width=0.5\columnwidth]{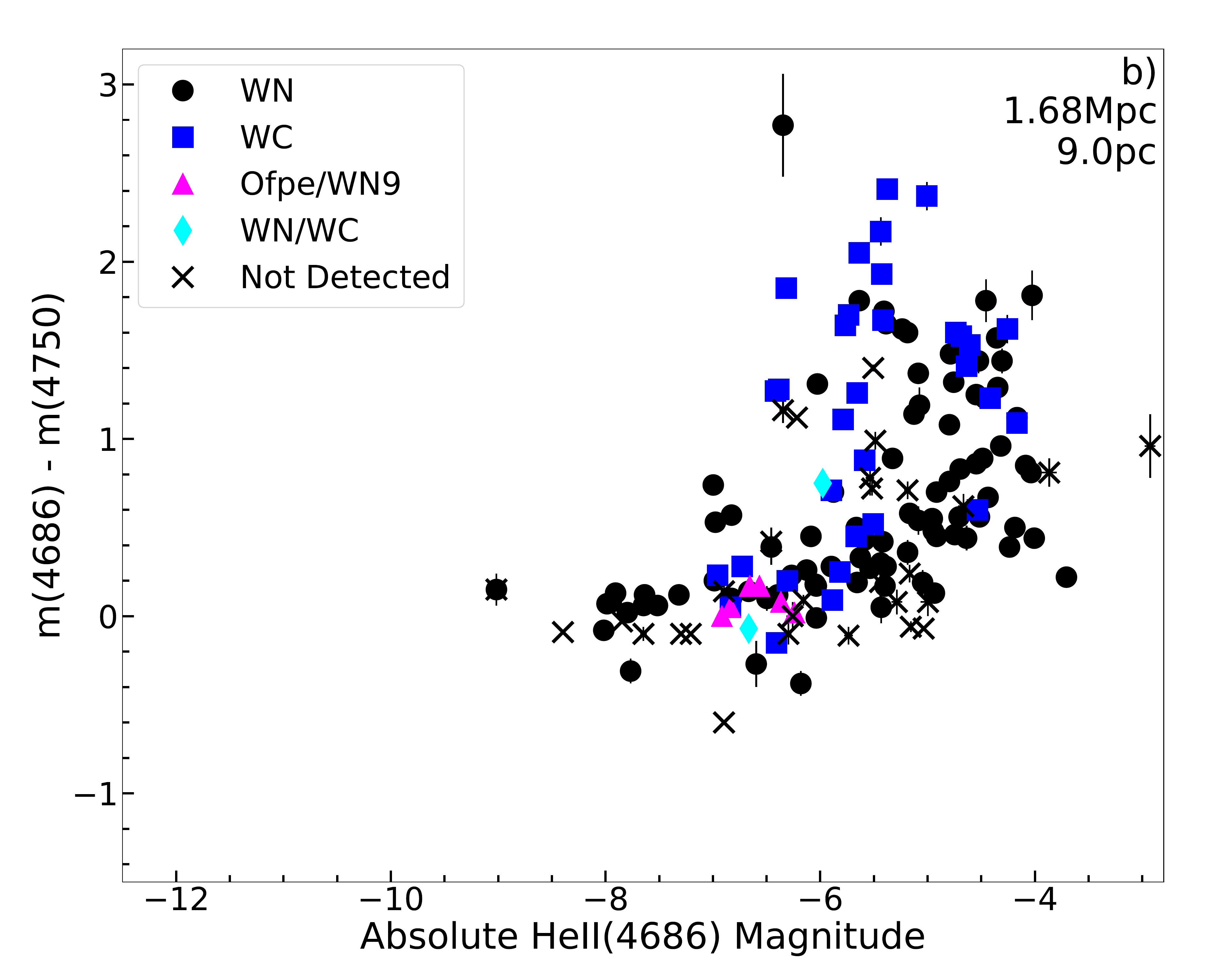} \label{2_deg}}
\vspace{-0.5cm}
\subfigure{\includegraphics[width=0.5\columnwidth]{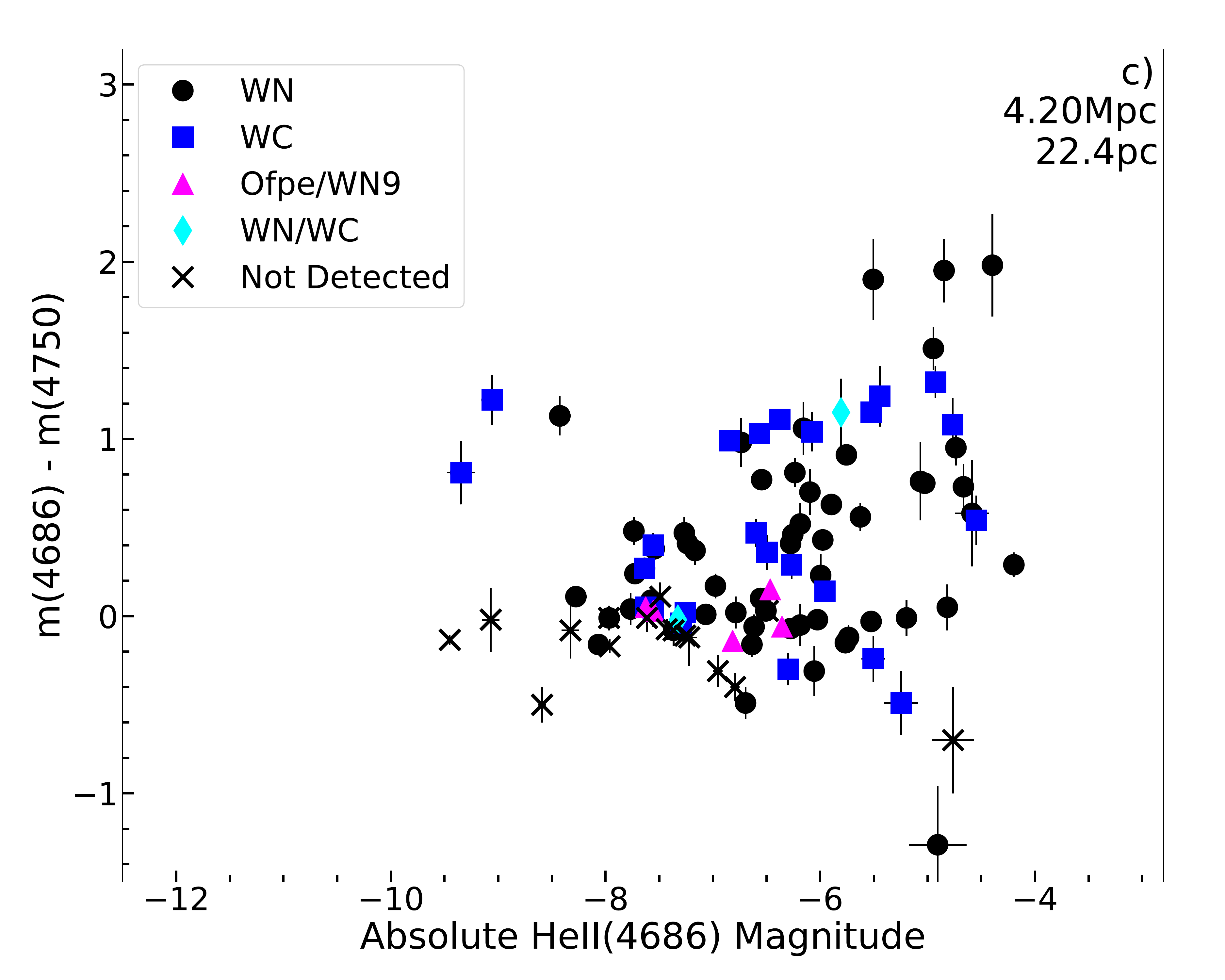} \label{5_deg}}
\subfigure{\includegraphics[width=0.5\columnwidth]{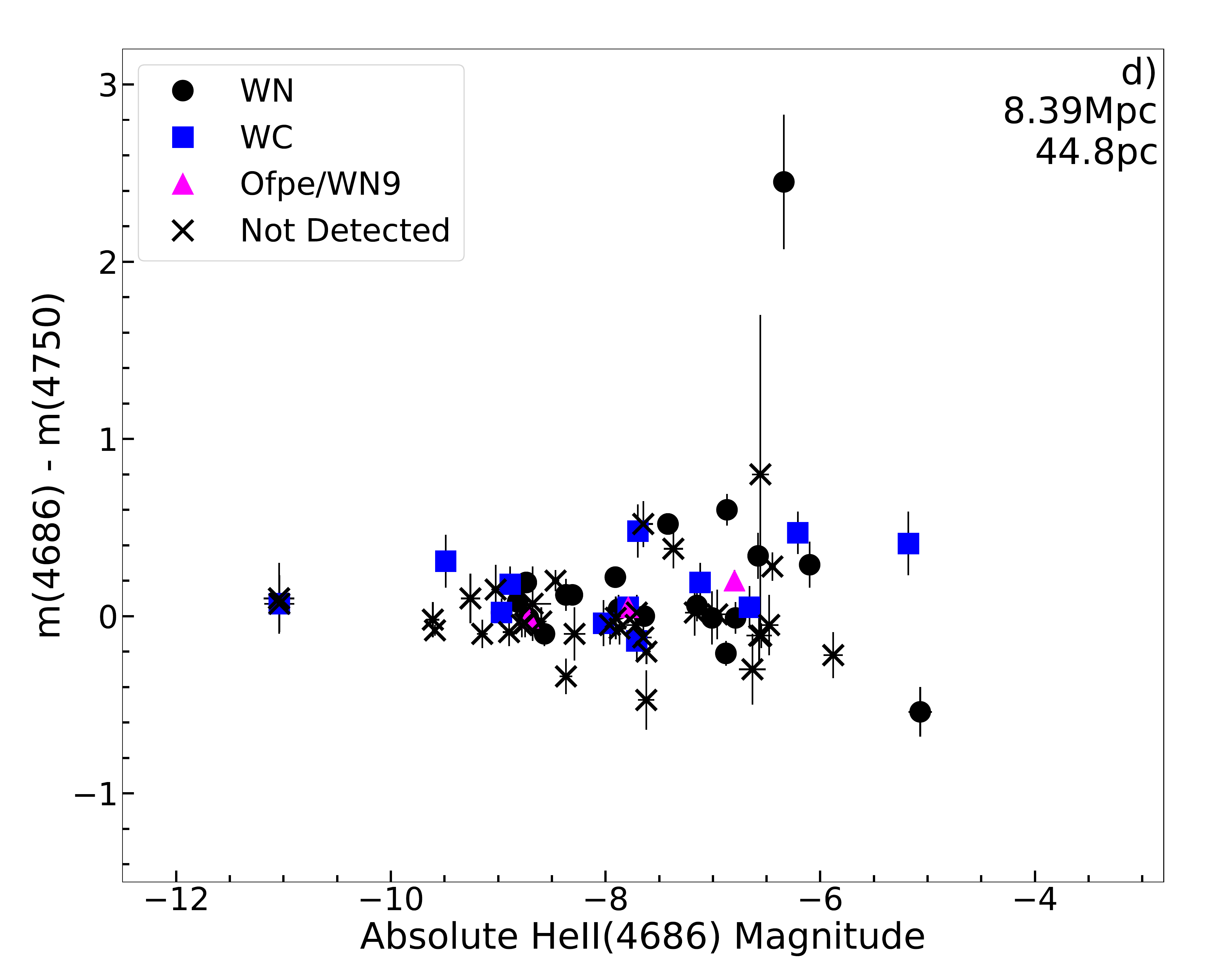} \label{10_deg}}
\subfigure{\includegraphics[width=0.5\columnwidth]{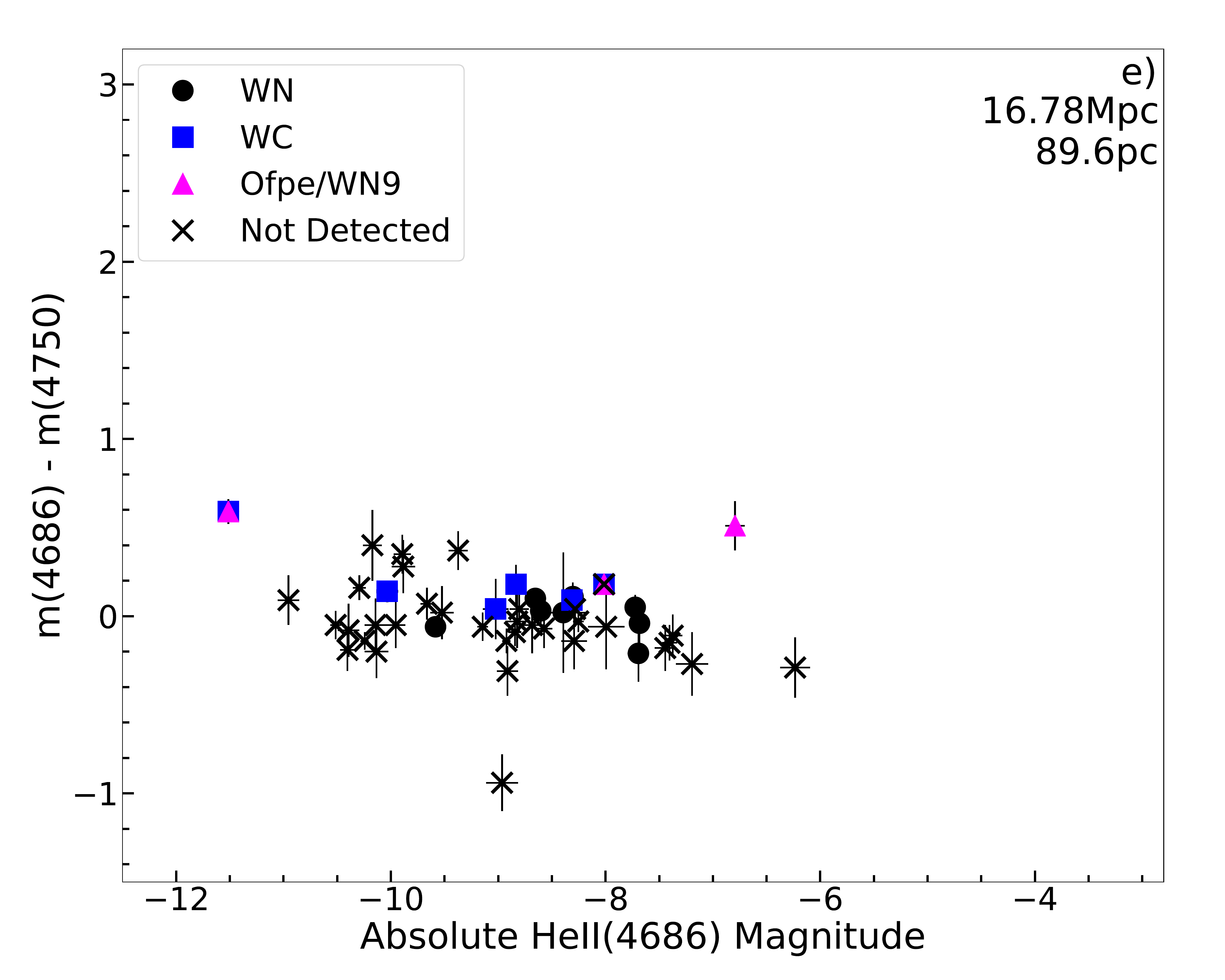} \label{20_deg}}
\subfigure{\includegraphics[width=0.5\columnwidth]{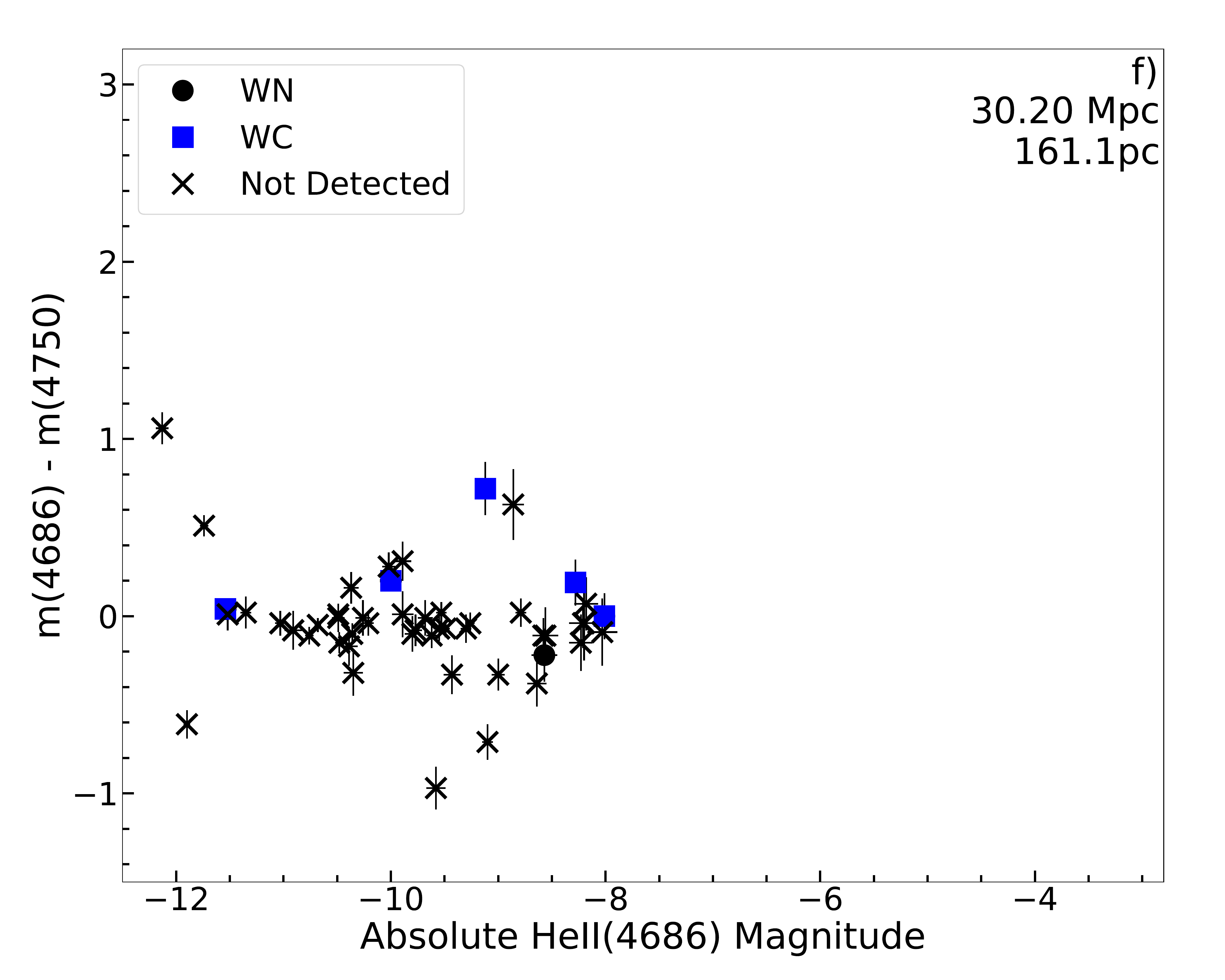} \label{36_deg}}
\vspace{-0.5cm}

\caption{Comparison of m$_{4686}$-m$_{4750}$ emission line excess versus
  M$_{4686}$ for the WR sources in M33 at different spatial
  resolutions. We note that depending on the WR environment the WR source plotted at each distance may be a single WR star, a binary WR system or an unresolved region containing multiple stars including WR stars; the latter becomes more common as distance increases. Different WR subtypes are highlighted and where a single WR source contains multiple WR stars of different subtypes we have over-plotted both subtypes. We identify binary WC and WN stars from \citet{Neugent2014} in a) only by open squares or open circles, respectively. Where possible, photometry (and errors) of the non-detections have been added for completeness using the a priori information from NM11. Many of these sources indicate an excess emission of around 0\,mag demonstrating the effect of line dilution at poorer spatial resolutions. All plots are on the same scale, highlighting the general trend of the WR sources, moving to increased brightness and weaker He\,{\sc ii} emission line excess as spatial resolution gets worse. These plots exclude WR candidates identified by eye for which no photometry was achieved.}
\label{mag_dist}
\end{figure}

Figure \ref{mag_dist} shows the magnitude and He\,{\sc ii} excess emission of the WR
stars at each spatial resolution. At high spatial resolutions we see a
range of magnitudes from M$_{4686}$\,=\,--3.5 to --\,9\,mag and emission
line excess strengths between 0.1\,--\,3\,mag. As the spatial resolution
decreases, the sources appear to get brighter and the emission line
excesses decrease. 

At $\sim$8\,Mpc ($\sim$45\,pc in spatial resolution) all the WR sources
have M$_{4686}$\,$<$\,--\,5\,mag and m$_{4686}$--m$_{4750}$\,$<$\,0.8\,mag (with the
exception of one source) (Figure \ref{mag_dist}d)). The emission line
dilution is so severe that many sources end up with
m$_{4686}$--m$_{4750}$$<$0 and consequently we only identify $\sim$40\%
of the WR stars. Many of the WR stars identified are done so
through inspection of the continuum subtracted image rather than
photometry. This is likely due to increased photometric errors and
increased line dilution as a result of crowding which subsequently
results in fewer 3$\sigma$ detections.

The farthest distance we investigate is 30\,Mpc, or an equivalent
spatial resolution of $\sim$160\,pc, where we detect only 15\% of WR
stars in the central pointing. The magnitude distribution, M$_{4686}$ versus
m$_{4686}$--m$_{4750}$ excess, of the WR stars is shown in Figure
\ref{mag_dist}f). Again, using a priori information, we can see that for the known WR stars that were not detected in our analysis, the majority have m$_{4686}$--m$_{4750}$\,$<$\,0\,mag, suggesting that they are not WR stars. This clearly demonstrates how WR stars can be hidden by surrounding stars when they cannot be resolved sufficiently. We note that there
are three sources with m$_{4686}$--m$_{4750}$ $>$0.5\,mag which we
would have expected to detect based on their photometry. However, both
sources, \#118 with m$_{4686}$--m$_{4750}$\,=\,1.06$\pm$0.09 and source
\#57 with m$_{4686}$--m$_{4750}$\,=\,0.63$\pm$0.20 show no He\,{\sc ii} excess visible
in the continuum subtracted image so were discounted as WR candidates
at this low resolution, following our candidate criteria. However,
source \#54 which has m$_{4686}$--m$_{4750}$\,=\,0.72$\pm$0.15 and a
visible excess emission in the continuum subtracted image should have
been identified via photometry and has been missed through human
error. 

We only detect 6 WR sources at 30\,Mpc, however a priori
information from NM11 tells us that these 6 regions
host 30 WR stars, suggesting an average of 5 WR stars per source at
this lowest resolution. Such information will be useful when
estimating the complete WR population of other galaxies. In addition, Figure \ref{mag_dist}f) shows that of the 6 unresolved sources detected at 30\,Mpc, 5 host WC stars and only one hosts WN stars but again using a priori information to investigate the contents of each source we see that there are actually more WN stars detected in total than WC stars; this is discussed further in Section \ref{WNL_detection}.

\vspace{-0.4cm}
\subsection{Significance Level of WR candidates}

One criterion for WR candidate selection via photometry was that the
$\Delta$m He\,{\sc ii} excess had a significance level greater than
3$\sigma$, i.e. the magnitude of the excess emission was at least 3
times the error on the excess. This discounted some bonafide WR stars
(for example that only had a 2$\sigma$ detection) but if a He\,{\sc ii} excess was
visible then these could be identified as candidates during inspection of the continuum subtracted image. However, if one wanted
to survey a galaxy without the manual, time-consuming image
inspection, what does our survey tell us about the expected
$\sigma$--value for increasing spatial resolutions?

\begin{figure}
    \centering
    \includegraphics[width=\columnwidth]{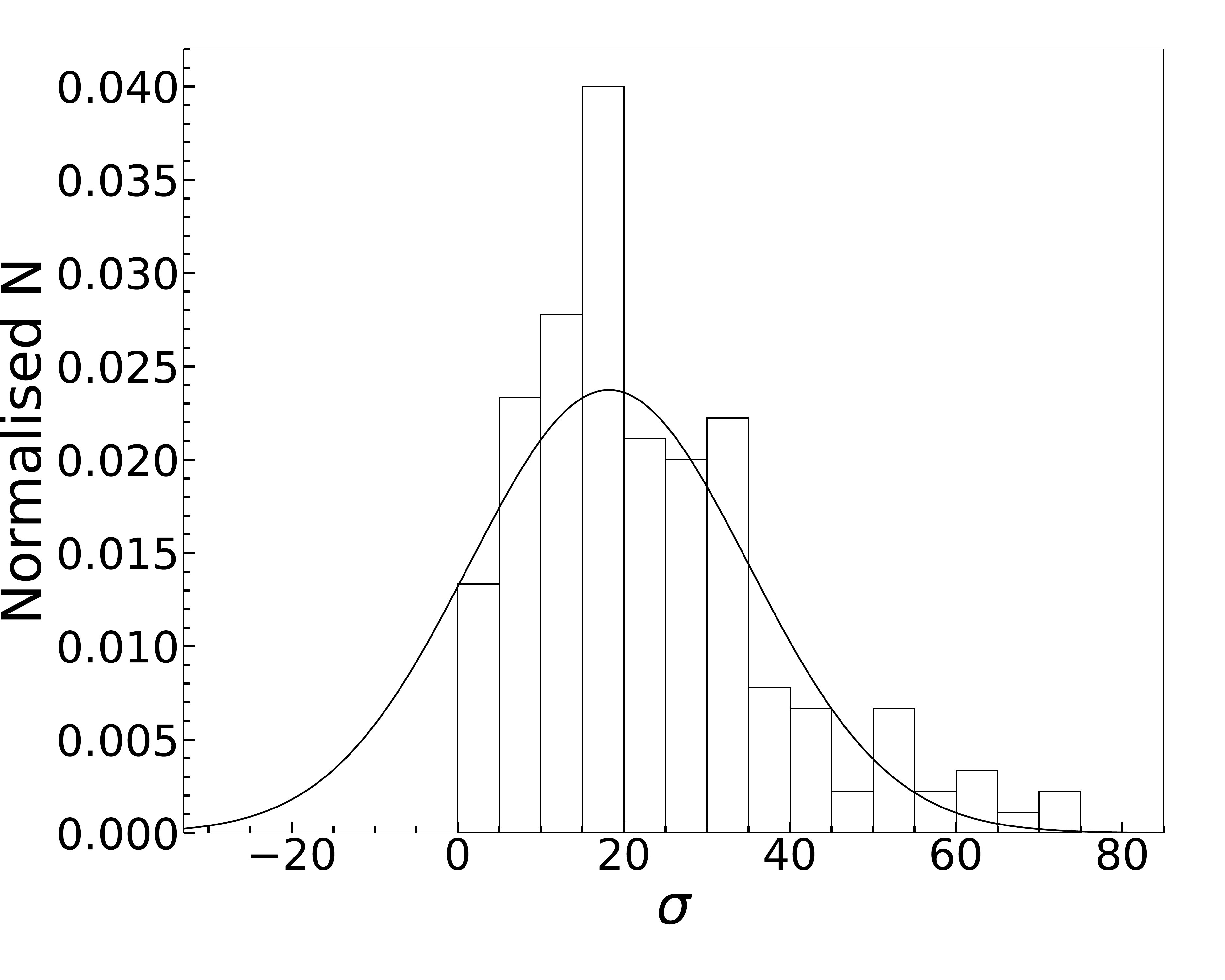}
   \vspace{-0.7cm}
    \caption{The distribution of the significance of each WR candidate He\,{\sc ii} excess emission ($\sigma$) for the original M33 image at 0.84\,Mpc. A Gaussian profile is fit to show the peak of the distribution at 18$\sigma$.}
    \label{sigma_plot_0deg}
\end{figure}

We plotted a histogram of the significance level for every WR
candidate identified, whether via photometry or from the continuum
subtracted image, and determined the peak of the distribution for each
spatial resolution by fitting a Gaussian profile. Figure \ref{sigma_plot_0deg} shows the distribution for the non-degraded image with a spatial resolution of $\sim$4.5\,pc. This peak represents the most likely significance
of the He\,{\sc ii} excess of a WR star at that spatial resolution; the
results for all resolutions are shown in Table \ref{sigma_limits}. It is clear that as the spatial resolution of the image gets poorer the significance of detections reduces. For the original (stacked) image the average WR
detection was at 18$\sigma$ falling to 8$\sigma$ at 22.4\,pc resolution
and below 3$\sigma$ for any resolution above $\sim$40\,pc. We note that
there are so few photometric measurements for WR stars in the 161\,pc
(d\,=\,30.2\,Mpc) that no meaningful value of $\sigma$ can be determined.

\onecolumn

\twocolumn

Our results demonstrate the impact that line dilution can have on detecting WR
candidates, decreasing the significance of the He\,{\sc ii} detection. We find an average 2$\sigma$ detection at 44.8\,pc which is consistent with the work of \citet{Sandford2013}. They investigate the WR population of NGC 6744
at 11.6\,Mpc using VLT FORS imaging of 0.7\,arcseconds, corresponding to
a spatial resolution of $\sim$40\,pc. They find that $\sim$40\% of their
candidates are detected with a significance level of 3$\sigma$,
$\sim$20\% with a 2$\sigma$ significance. An additional 40\% are only
detected in the He\,{\sc ii} image so an excess cannot be
determined. Whilst it is possible that their sample is contaminated
and not all of their candidates are bonafide WR stars, their and our work both
suggest that a 3$\sigma$ significance level for the detections of WR
candidates may be too stringent for ground-based WR surveys beyond
$\sim$10\,Mpc. Relaxing the sigma level brings other problems, such as more false positives so a more in depth spectroscopic study of 2$\sigma$ WR candidates is needed to fully understand if WR stars can be detected efficiently at this level. 

\begin{table}
\centering
\caption{The peak significance level for a WR source at each
  spatial resolution. We require a $>$3$\sigma$ significance for
  a source to be classed as a WR candidate. At 30\,Mpc there are so few
  sources detected that the peak of the distribution cannot be
  determined.}
\begin{tabular}{cc}
\hline
Resolution (pc) & Peak $\sigma$ \\
\hline
4.5 & 18 \\
9.0 & 15 \\
22.4 & 8 \\
44.8 & 2 \\
89.6 & 1.5 \\
161.1 & -- \\
\hline

\end{tabular}
\label{sigma_limits}
\end{table}

\subsection{Detection limits of images}

In Table \ref{detections} we note that only 15\% of the WR stars in
the M33 image degraded to $\sim$160\,pc are detected. Figure
\ref{mag_dist} f) shows that (most of) the non--detections are due to
increased errors and the effect of line dilution decreasing the
emission-line excess. However, as we degrade the original image we also increase the
background noise causing some of the sources to be indistinguishable
from the background. We must ask, are the detection limits of the
degraded image sufficient to detect all of the WR population, even if
we can't detect them as individual stars?

By considering all of the sources detected in the M33 images,
irrespective of whether they are WR stars, we determine the detection
limit of our images. The detection limit of each degraded image is
calculated from a histogram of all sources detected in the image, not
just WR stars. The 100\% completeness limit is noted where the
distribution peaks, although as expected some sources are detected at
fainter magnitudes (see \citet{Bibby2010} for details). Table
\ref{detection_limits} reveals that the 100\% detection limit of the
original image for all stars (including non--WR stars) is
M$_{4686}$\,=\,--2.3\,mag, increasing in absolute brightness to
M$_{4686}$\,=\,--9.75\,mag at 30.2\,Mpc. Looking at the magnitude distribution
of detected WR stars at 30.2\,Mpc (Figure \ref{mag_dist} f)), $\sim$50\%
of the known WR stars are fainter than this detection limit.

Although we know that at poorer spatial resolution M$_{4686}$
increases in absolute brightness as sources blend, this detection
limit at 30.2\,Mpc could mean that we are missing WR stars as a result of
our images not being deep enough. To quantify these detection
statistics we inspected the 30.2\,Mpc image and photometry files and
concluded that even though only 30 (15\%) of the WR stars were
identified as WR candidates, 87\% of the total WR population was in
fact visible in the image, albeit in crowded, unresolved regions with
no He\,{\sc ii} excess. The remaining 13\% of WR stars were hidden by
the background noise of the image and could potentially be
distinguished with increased S/N. This suggests that at most, with
increased exposure times, we could detect and identify $\sim$28\% (15\%$+$13\%) of the WR population at a spatial resolution of 160\,pc.

We note that a detection limit of M$_{4686}$\,=\,--9.75\,mag for the
magnitude distribution of WR stars at 0.84, 1.69 and 4.2\,Mpc (4--20\,pc
resolution)(Figure \ref{mag_dist} a)-c)) would not be sufficient to
detect \textit{any} of the known WR stars. All the WR sources detected
at 30.2\,Mpc host multiple WR stars.

\begin{table}
\centering
\caption{The 100\% completeness limit as a function of
  distance. Nearby, we are able to detect fainter objects down to
  M$_{4686}$\,=\,--2.3\,mag. At larger distances these fainter objects
  are missed due to blending and increased noise.}
\begin{tabular}{cc}
\hline
Distance (Mpc) & M$_{4686}$ \\
\hline
0.84 & --2.3 \\
1.68 & --2.75 \\
4.20 & --4.5 \\
8.39 & --6.75 \\
16.78 & --8.75 \\
30.20 & --9.75 \\
\hline

\end{tabular}
\label{detection_limits}
\end{table}

\subsection{Detecting Different WR Subtypes}
\label{WNL_detection}

So far we have mainly looked at WR detectability of WN and WC stars. In this section we look in more detail about what this
resolution study tells us about the detectability and identification
of different WN and WC sub-types.

Based on the stronger emission lines of WC stars, most WR surveys
assume that they are more complete for WC stars than WN stars. To test
this we use the known subtype of our detected WR stars (from NM11 and references therein) and we use the classification scheme of \citet{Smith1968} to categorise them as early--WN (WNE) and late--WN (WNL). Table \ref{nwn_nwc} shows the number of each subtype detected at each spatial resolution; those simply noted as ``WN'' do not have any detailed classification in NM11. We do
not split WC stars into early and late types because there are only two
WCL stars in M33. Transition WN/WC and Ofpe/WNL stars are also listed
for completeness but low number statistics makes any conclusions
unreliable. 

Using the a priori information from NM11 for the WR subtype we can see that in the stacked but undegraded image we detect 98\% of the WC stars, 100\% of the WNE stars and 90\% of the WNL stars. This decreases by $\sim$15\%  for each subtype at a spatial resolution of 9\,pc. At our largest spatial resolution of 161\,pc we detect only 14\% and 8\% of the WC and WNE stars, respectively but are able to recover 40\% of the WNL stars.

\begin{table}
\centering
\caption{The number of detected sources by subtype at the increasing
  distances/spatial resolutions. WN stars are those not identified as
  either WNL or WNE in NM11 and WN/WC are transition
  stars as defined by \citet{Conti1989}.}
\begin{tabular}{c@{\hspace{2mm}}c@{\hspace{2mm}}c@{\hspace{2mm}}c@{\hspace{2mm}}c@{\hspace{2mm}}c@{\hspace{2mm}}c@{\hspace{2mm}}c}
\hline
Distance & Spatial & WN & WNE & WNL & WC & WN/ & Ofpe/  \\
(Mpc) & Res. (pc) &    &     &     &    &  WC  & WNL    \\
\hline
 0.84$^{1}$    & 4.5                     & 28 & 75  & 31  & 51 & 2     & 9         \\
\hline 
0.84           & 4.5                     & 20 & 75  & 28  & 50 & 2     & 8         \\
1.68           & 9.0                     & 18 & 65  & 23  & 43 & 2     & 7         \\
4.20           & 22.4                    & 11 & 49  & 23  & 41 & 2     & 6         \\
8.39           & 44.8                    & 3  & 29  & 15  & 25 & 1     & 3         \\
16.78          & 89.6                    & 2  & 14  & 14  & 14 & 0     & 4          \\
30.20          & 161.1                   & 2  & 6   & 12  & 7  & 0     & 3          \\
\hline 
\multicolumn{8}{l}{$^{1}$ In the central pointing of M33 as observed by NM11, see}.\\
\multicolumn{8}{l}{Section \ref{M33} for details.} \\
\end{tabular}
\label{nwn_nwc}
\end{table}

One explanation is that WNL stars are intrinsically brighter than
other WR subtypes \citep{Sander2012}. However, updated distances from Gaia DR2 has revealed that a correlation between absolute visual magnitude and WR subtype is weak \citep{Hamann2019, Rate2020}. Moreover, NM11 investigate the average M$_{V}$ for each
individual subtype in the original image and find no distinct differences and a large standard deviation for each subtype. Overall this suggests that the brightness is not responsible for the increased detection of WNL stars at greater distances.

\begin{figure}

\includegraphics[scale=0.98, trim={2cm 13cm 2cm 0cm}, clip]{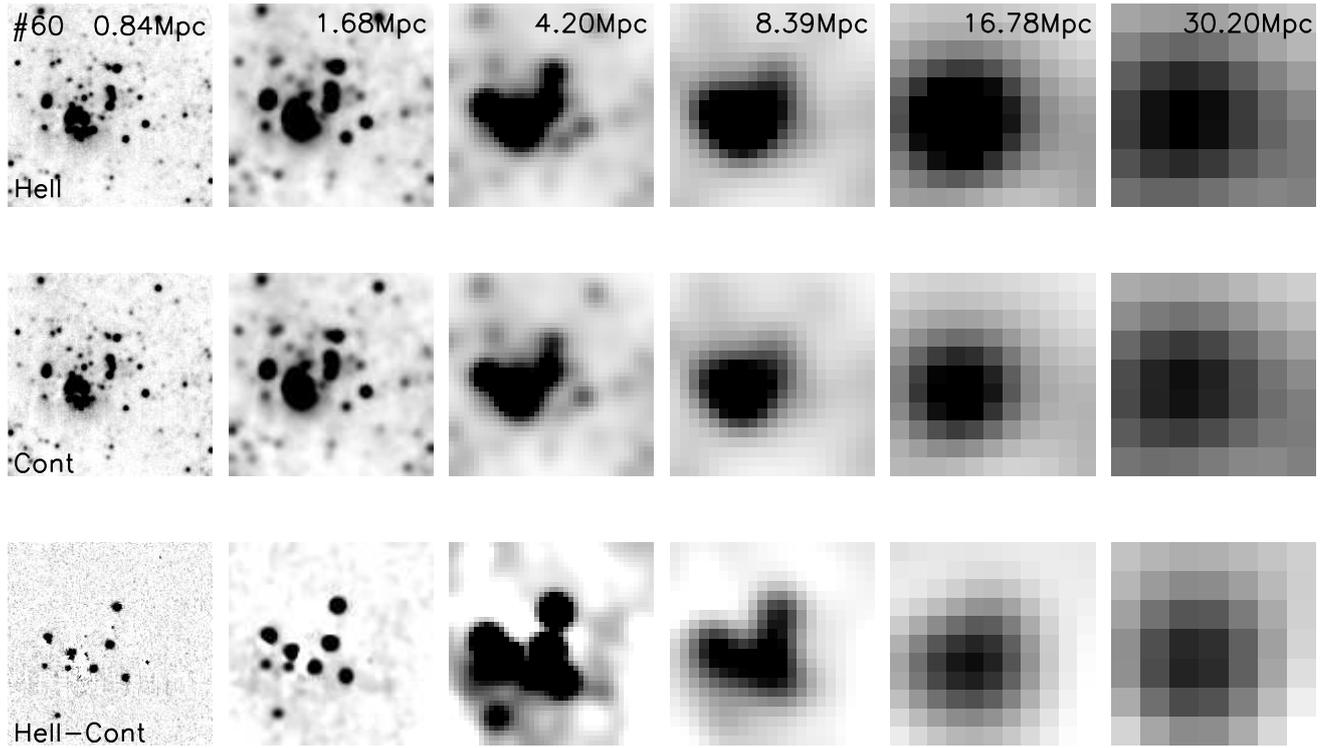}

\caption{NGC 595 in M33 containing 11 WR stars,
  including 9 WNL stars, 1 WC and 1 WN star. This figure shows the
  He\,{\sc ii} (top), continuum (middle) and continuum-subtracted
  (bottom) image for a $\sim$ 1 arcminute box around the region. The
  image is degraded from 0.84\,Mpc to 30.2\,Mpc (left to right), showing
  how the individual WR sources become part of the a single source at
  higher distances. The He\,{\sc ii} excess (shown as dark emission in the bottom row) can be detected out to 30.2\,Mpc.}

\label{WNL_cluster}

\end{figure}

Another possible explanation as to why WNL stars appear to be easier
to detect with increasing distance is that they are more concentrated in terms of their spatial
distribution across the M33 galaxy compared to other subtypes. This
would mean that their emission line excesses, though weak, combine to
produce a stronger line that can still be detected at large
distances. All of the 12 WNL subtypes detected at an equivalent
distance of 30.2\,Mpc are located within just two unresolved regions. Figure \ref{WNL_cluster} shows the H\,{\sc ii} region NGC 595 in M33 which hosts 9 of the 12 WNL stars
detected at 30.2\,Mpc. At 0.84\,Mpc the region is resolved and the He\,{\sc ii} excess
for each star can be quantified, however at 30.2\,Mpc the unresolved
nature of the region means all WR stars are still detected, albeit in a single region with one He\,{\sc ii} emission line excess representing all
the WR stars. Consequently, despite the emission lines of WNL stars
being weaker than WNE and WC stars, their spatial distribution in this
region means the He\,{\sc ii} excess is still detectable above the
continuum making WNL stars more detectable than one would expect at
such distances.

Our survey shows that this might indeed be unique to WNL stars. The
seven WC stars detected at 30.2\,Mpc lie in five unresolved regions, two
of which contain no other WR stars. Whilst it is not a surprise that
single WC stars can be detected at larger distances given their strong
emission line excess, it does suggest WNL stars may have a preference
to be in more dense regions/clusters than WC stars. 

\citet{Smith2015} find that Luminous Blue Variables (LBV) in the LMC are more isolated than O
stars and WR stars. The explanation for this is that LBVs must result from binary evolution of lower mass stars that have had more time to migrate further from their natal environment. In this work they find also that WR stars are more dispersed than O stars again having had even a short time to move away from their natal environment, albeit not as far as LBVs. They find that WC stars are more spatially dispersed from O stars than WN stars, which is consistent with our ability to detect WNL stars in our survey. However \citet{Smith2015} interpret
this as WC stars being older than WN stars, rather than more massive
(younger) and they do not investigate whether the different distributions of WN stars
relative to WC stars is statistically significant. Further investigation into the spatial
distribution of WR subtypes is required to see if any clumping of
individual subtypes is truly present and subsequently if this leads to a higher detection rate for a specific subtypes e.g. WNL stars as this work suggests.

\onecolumn

\twocolumn

\section{Discussion}
\label{discussion}

\subsection{Comparisons with stellar evolutionary models}

\begin{figure}
\includegraphics[trim={0cm 0cm 6cm 12cm}, clip, width=1.05\columnwidth]{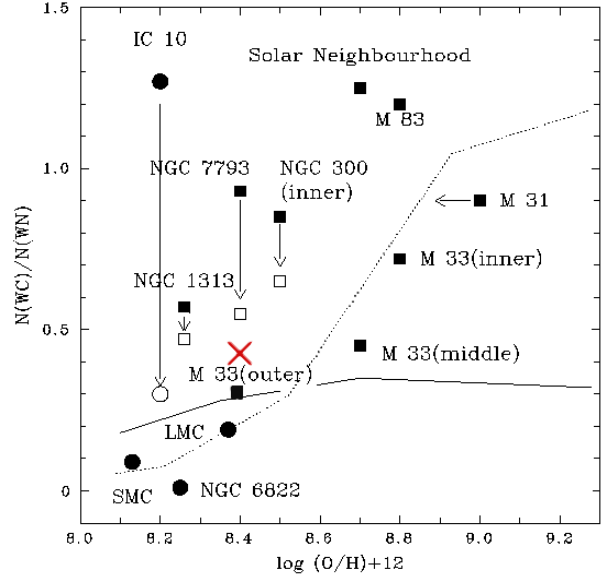}

\caption{Adapted from \citet{Bibby2010} (their figure 13a)) showing
  the WC/WN ratio in a number of galaxies at different metallicities. The
  observed ratios are plotted as solid points and the ratio corrected
  for WC bias and detection limits are the open points. The WC/WN ratio
  for NGC 7793 is additionally corrected for spatial resolution and line dilution using this work and is indicated by the red X. The result for NGC 7793 is more in line with evolutionary
  predictions from \citet{Eldridge2006}(dotted line) and
  \citet{Meynet2005} (solid line).}
\label{WC_WN_plot}
\end{figure}

WR stars are produced by the removal of the outer envelope of a star by metal--driven stellar winds.  Consequently, more WR stars are found in metal--rich regions because the stellar winds are more efficient at stripping the stars. This causes the lower mass limit for a WR star to decrease, with some stars that would normally become Red Supergiants now becoming WN stars. Similarly, the mass limit for a WC star decreases so the WC/WN ratio changes with metallicity \citep{Eldridge2006}. However, direct comparison of observational results with
stellar evolutionary models has revealed a higher observed WC/WN ratio
than predicted across all metallicities. Most studies (e.g. NM11,
\citet{Bibby2012}, \citet{Crowther2003}) correct for completeness in
terms of an absolute magnitude limit and an observational bias towards
WC stars to account for the discrepancy. Figure \ref{WC_WN_plot},
adapted from \citet{Bibby2010} shows that correcting for completeness in this way reduces the
observed ratio towards the predicted values (solid points to open points) but
theoretical predictions are still not in agreement with observations. 

Most recently, \citet{Stanway2020} use the WC/WN ratio to investigate the binary fraction of the stellar population and whether this can be recovered from observations of stellar populations. They find that the WC/WN ratio decreases sharply with increasing binary fraction but note that number ratios from unresolved stellar populations must be used with caution when comparing to stellar models. 

If the detectability of WN and WC stars is not equally affected by line dilution then this will have an impact on measured WC/WN ratios and any conclusions from comparison with stellar evolutionary models.
The work presented here may be able to provide a further correction accounting for spatial resolution. For example, if we consider the WR analysis of NGC 7793 for which the relevant data is available, \citet{Bibby2010} observe a WC/WN ratio of $\sim$0.95, decreasing to $\sim$0.5 after applying standard completeness corrections. This is still not in agreement with stellar evolutionary models from which we expect the a ratio of $\sim$0.2--0.3. The resolution of the VLT/FORS1 NGC 7793 data is 1.3 arcseconds, corresponding to a physical scale of $\sim$25pc assuming a distance of 3.91\,Mpc. From our analysis of M33 (see Table \ref{nwn_nwc}) if we exclude transition WN/WC and Ofpe/WNL stars, we see that the WC/WN ratio at a resolution of 22.4\,pc (distance of 4.2\,Mpc) is 0.49 whereas at the best spatial resolution of 4.5\,pc we find WC/WN = 0.38. If this 0.11 correction is applied to the NGC 7793 data then this suggests (see red X on Figure \ref{WC_WN_plot}) that about half of the discrepancy between theoretical predictions and observations can be accounted for by WR stars not detected due to line dilution (rather than due to magnitude limits).

\subsection{Implications for supernova progenitor detections}

\begin{figure}

\includegraphics[width=\columnwidth, trim={2cm 12.5cm 2.5cm 2cm}, clip]{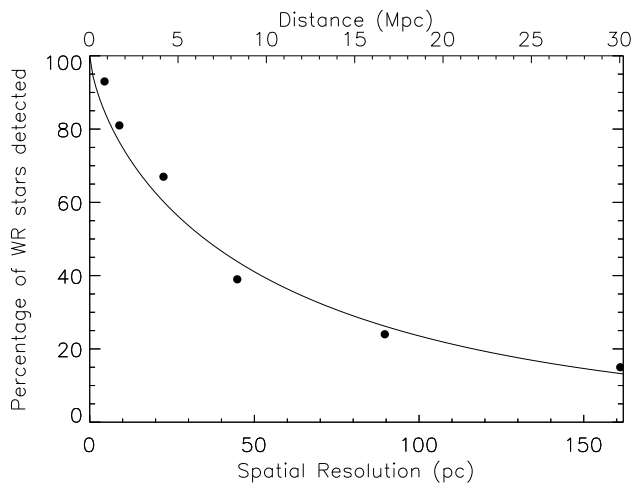}

\caption{Results of the image degradation showing the percentage of WR
  stars detected at increasing distance and decreasing spatial
  resolution as listed in Table \ref{detections}.}
\label{results_fig}
\end{figure}

Figure \ref{results_fig} shows the percentage of WR stars that you would
expect to detect as a function of spatial resolution and distance, resulting from our analysis. The WR stars in this figure are those detected at that spatial resolution via photometry or inspection of the image, as outlined in Section \ref{method}. Each WR that is detected with a He\,{\sc ii} excess is included, even if there are multiple WR stars within one unresolved source. 

These data are empirically fit by equation \ref{equation} which constrains the fit to 100\% at a spatial resolution of 0\,pc, and is applied up to a spatial resolution of 160\,pc. This equation, plus the scatter about it seen in Figure \ref{results_fig}, can be used to give an estimate of completeness as a function of spatial resolution which is useful when planning observations. 

\begin{equation}
\centering
    y = 100 \times 10^{-0.025x^{0.7}}
 \label{equation}
\end{equation}

It is clear that beyond a spatial resolution of $\sim$20pc, typical of a
classical H\,{\sc ii} region \citep{Conti2008}, we cannot detect a
sufficient ($>$60\%) number of WR stars to consider a survey
complete. At ground-based resolution of $\sim$1 arcsecond, 20\,pc corresponds to $\sim$4\,Mpc,
which limits the number of galaxies we can survey to a reasonable degree of completeness. This distance limit drastically reduces the
probability of detecting a pre--SN WR star; there have been no Type Ibc
ccSN within $\sim$4\,Mpc within the last 10 years, and only 4 since on
record in total (SN1954A, SN1962L, SN1983N, SN2002ap and SN2008dv).
\citep{Guillochon2017}).

However, if HST or Adaptive Optics is utilised for WR surveys then the 0.1 arcsecond
resolution affords a factor of $\sim$10 increase in distance, out to
40\,Mpc.  Within this distance, 138 Type Ibc SN have been recorded
\citep{Guillochon2017,Barbon2008} which is a significant increase on
4\,Mpc and would increase our chances of identifying a pre--SN WR star if
such WR surveys existed.

From Table \ref{detections} it is clear that visual inspection of the image results in a more complete WR survey. Whilst \citet{Morello2018} successfully use a machine-learning approach to identify new WR stars in the Milky Way, it is unclear if such an approach can be successfully employed for more distant galaxies where we see changing photometric properties with changing spatial resolution (recall Figure \ref{mag_dist}).

\subsection{Implications for current and future surveys}

Here we consider how detectable WR stars are at large distances based on both spatial resolution and magnitude, accounting for current and upcoming instrumentation. We set a threshold of M$_V$\,=\,--3\,mag for the faintest WR stars which is consistent with the analysis of WR stars in the LMC, SMC and M33 by \citet{Neugent2011} and determine m$_V$ for 7 distances up to 60\,Mpc. Based on our previous experience we assume an exposure time of 1 hour and use the narrow-band VLT/FORS2 sideband filter, He\,{\sc ii}/6500$+$49 \footnote{http://www.eso.org/sci/facilities/paranal/instruments/fors/doc/ VLT-MAN-ESO-13100-1543\textunderscore{P06}.pdf} centred on 4781\AA~ with the standard resolution collimator to calculate the S/N ratio achievable in the continuum for the faintest WR stars. We also look at VLT/MUSE IFU in Wide Field Mode (WFM) with extended spectral range to include He\,{\sc ii} $\lambda$4686 line. The Muse User Manual\footnote{http://www.eso.org/sci/facilities/paranal/instruments/muse/
doc/ESO-261650\textunderscore{MUSE}\textunderscore{User}\textunderscore{Manual}.pdf} gives the relevant parameters for no AO WFM with 0.2\arcsec per pixel sampling. Overall the read out and dark count contribution to the noise is small and the noise is sky-dominated under grey observing conditions. Sky values for the VLT are taken from \citet{Noll2012} and are assumed to be similar at the ELT site.  

We assume a nominal resolution of 0.8\arcsec~ for VLT and 0.06\arcsec~ for ELT based on a 20mas sampling as suggested by specification document for ELT-IFU observations in the visible range.
\footnote{https://www.eso.org/sci/facilities/eelt/docs/ESO-191883\textunderscore{2}\textunderscore{Top}\textunderscore{Level}\textunderscore{Requirements}\textunderscore{for}\textunderscore{ELT}-IFU.pdf} The throughput of the IFU in the visible range is the least certain aspect of our analysis which we take as $\sim$30\% \footnote{www2.physics.ox.ac.uk/research/visible-and-infrared-instruments/harmoni} compared to $\sim$60\% for VLT/FORS2 with the He\,{\sc ii} filter and $\sim$17\% for VLT/MUSE.

The results of our analysis are presented in Table \ref{VLT_ELT}. 
Using VLT with FORS2 or MUSE without adaptive optics we expect to be able to detect the faintest (M$_{V}$\,=\,--3) WR 
stars out to a distance of $\sim$4.5\,Mpc in terms of S/N. However, at this distance the spatial resolution of $\sim$16\,pc would likely hinder our ability to detect and resolve all WR stars as indicated by Figure \ref{results_fig}. This is consistent with previous WR surveys at similar distances (\citealt{Hadfield2005,Bibby2010,Bibby2012}). Some WR and WR star complexes have been detected with MUSE (e.g. in NGC300 at 1.9 Mpc, \citep{Roth2018}; in NGC4038/39 at 18 Mpc, \citep{Gomez2021}.) Looking to future instrumentation, namely ELT/HARMONI with adaptive optics available at visible wavelengths, we expect to be able to achieve sufficient S/N to detect many individual WR stars out to 30\,Mpc. Such future observations would allow for studies of WR populations in many different types of galaxies, in different environments.

\begin{table*}
    \centering
    \caption{Single WR continuum detectability with spatial resolutions and signal-to-noise values achievable for observations of WR stars up to $\sim$60\,Mpc. We present data for observations for VLT/FORS2 without AO, VLT/MUSE without AO in extended mode and ELT/HARMONI with AO. FORS2 and MUSE have a resolution of 0.8\arcsec whilst HARMONI has a resolution of 0.06\arcsec. We assume M$_V$\,=\,--3\,mag for the faintest WR stars and grey sky background.}
    \begin{tabular}{c@{\hspace{2.5mm}}c@{\hspace{2.5mm}}cc@{\hspace{2.5mm}}c@{\hspace{2.5mm}}c@{\hspace{2.5mm}}c@{\hspace{2.5mm}}ccc}
    \hline 
    Scale        &  Distance    & Example     & m -- M  & m(V)     & S/N     & S/N &  0.8\arcsec~ Resolution  &  S/N     & 0.06\arcsec~  Resolution   \\
    (pc/\arcsec) & (Mpc)        & Environment & (mag)   & (mag)    & FORS2 & MUSE & (pc/resolution)  &  HARMONI    &   (pc/resolution) \\
    & & & & & (no AO) & (no AO - E) &  & (with AO) \\
    \hline
    4.07        & 0.84    & M33            & 24.62     & 21.62    & 170.90    & 90.93  & 3.26    & 1355.99   & 0.24 \\
    8.14        & 1.68    &                & 26.13     & 23.13    & 45.80     & 24.37  & 6.52    & 652.81    & 0.49 \\
    20.36       & 4.20    & Sculptor Group & 28.12     & 25.12    & 7.51      & 3.99   & 16.29   & 213.38    & 1.22 \\
    40.68       & 8.39    &                & 29.62     & 26.62    & 1.89      & 1.00   & 32.54   & 72.12     & 2.44 \\
    81.35       & 16.78   & Fornax Cluster & 31.12     & 28.12    & 0.47      & 0.25   & 65.08   & 20.26     & 4.88 \\
    146.41      & 30.20   &                & 32.40     & 29.40    & 0.15      & 0.08   & 117.13  & 6.45      & 8.79 \\
    292.83      & 60.4    & Hydra Cluster  & 33.91     & 30.91    & 0.04      & 0.020  & 234.26  & 1.63      & 17.57 \\
    \hline 
    \end{tabular}
    
    \label{VLT_ELT}
\end{table*}

\section{Conclusions}
We present a WR survey of M33 using narrow-band images degraded to 5
different absolute spatial resolutions, representative of distances of up to
30\,Mpc for 1.1 arcsecond apparent resolution. We find that: 

1) Emission line dilution resulting from poorer spatial
resolution drastically reduces the detectability of WR stars out to 30\,Mpc.

2) Based on the distribution of the significance level of the He\,{\sc ii} excess for each WR candidate detection we conclude that inspection of the image by eye is vital if we are to detect as many WR stars as possible.

3) Resolution plays more of a role in limiting the detection of WR stars than S/N.

4) WNL stars appear to be more clustered compared to WC stars,
suggesting that WC stars are older than WNL stars, rather than more
massive.

5) If a correction for line dilution as a result of spatial resolution
is applied, WC/WN ratios are more in line with predictions from
stellar evolutionary models.

In summary, the emission line dilution that occurs as a result of
being unable to resolve bonafide WR stars from their neighbours
severely impacts our ability to identify these stars. This has a wider
impact on identifying supernova progenitors as well as testing
predictions from stellar evolutionary models. 

Narrow-band imaging and spectroscopic surveys of galaxies expected to
have a significant WR population (e.g. NGC 6946, M83, M51) have
already been undertaken with ground-based facilities. These surveys
are typically limited to $\sim$10\,Mpc for the reasons quantified in
this paper. M101, at 6.5\,Mpc, is the only complete grand spiral galaxy
to be imaged with HST/WFPC3 using F469N narrow-band filters to isolate the He\,{\sc ii} emission. HST affords a similar
spatial resolution to the undegraded M33 imaging of $\sim$3\,pc and
reveals more WR candidates than would be expected from similar
ground-based surveys \citep{Shara2013}. 

We show that current instrumentation, e.g.VLT/FORS2 or MUSE limits WR surveys to within 4\,Mpc and that currently only narrow-band imaging with HST/WFC3 can produce the superior spatial resolution required to detect a significant number of WR stars out to larger distances. We investigate the ability of planned instruments such as ELT/HARMONI, with AO available at optical wavelengths and conclude that such instruments will significantly improve our ability to detect WR stars and allow us to undertake WR surveys of galaxies at distances of up to 30\,Mpc.

\section*{Acknowledgements}

The authors thank Philip Massey and Kathryn Neugent for providing them with the fully reduced narrow-band imaging of M33. We thank the referee, Paul Crowther, for providing helpful comments that improved this paper. We acknowledge
Aaron Brocklebank for the initial degrading of the images with
financial support from STFC. AJS acknowledges financial support from
the the UCLan Undergraduate Research Internship Programme 2019. We
also thank Molly Hawkin from Cardinal Newman College for her work on
Figure \ref{blink} and \ref{WNL_cluster} as part of the Ogden Trust
outreach program. This paper makes use of observations obtained with
the University of Central Lancashire's Moses Holden Telescope.

\section*{Data Availability}
The data underlying this article will be shared on reasonable request to the corresponding author.




\bibliographystyle{mnras}
\bibliography{m33_deg} 










\bsp	
\label{lastpage}
\end{document}